\documentstyle[11pt,fleqn]{article}
\def\ref{par\noindent\hangindent=6mm\hangafter=1}
\begin{document}
\vbox{
\rightline{IFUG-94/23}
\rightline{hepth/9411026}
\rightline{Phys. Rev. E 51, 5112 (May 1995) [BR]}
}
\baselineskip 8mm
%\draft
%\preprint{IFUG-94/23}
\begin{center}
{\bf Supersymmetric time-continuous discrete random walks}

\bigskip

Haret C. Rosu\cite{byline}

{\it Instituto de F\'{\i}sica de la Universidad de Guanajuato, Apdo Postal
E-143, L\'eon, Guanajuato, M\'exico}

Marco Reyes\cite{byline}

{\it Departamento de F\'{\i}sica, Centro de Investigaci\'on y Estudios
Avanzados del Instituto Polit\'ecnico Nacional, Apdo Postal 14-740,
M\'exico Distrito Federal, M\'exico}

\end{center}

\bigskip
\bigskip

\begin{abstract}

We apply the supersymmetric procedure to one-step random walks in one
dimension at the level of the usual master equation, extending a study
initiated by H.R. Jauslin [Phys. Rev. A {\bf 41}, 3407 (1990)]. A
discussion of the supersymmetric technique for this discrete case is
presented by introducing a formal second-order discrete master
derivative and its ``square root", and we solve completely,
and in matrix form, the cases of homogeneous random walks
(constant jumping rates).
A simple generalization of Jauslin's results to two uncorrelated axes
is also provided.
There may be many applications, especially to bistable and multistable
one-step processes.

\end{abstract}
\bigskip
PACS number(s): 05.40.+j, 11.30.Pb

\newpage
%%%%%%%%%%%%%%%%%%%%%%%%%%%  THE PAPER  %%%%%%%%%%%%%%%%%%%%%%%%%%%%%%%%
\section*{I. Introduction}

A number of interesting results have been obtained in the past for the
bistable Fokker-Planck equation \cite{BB} by means of the Witten
supersymmetric
approach \cite{W1}. Several years ago, in a paper hereafter denoted as I,
Jauslin \cite{J1} developed the method of supersymmetric
partners for time-continuous uncorrelated discrete random walks (RW's) in one
dimension (1D). He discussed briefly the factorization of the discrete master
operator, some of the properties of the superpartners, and as an application,
the so called ``addition of eigenvalues" method, generating bistability and
multistability, in the supersymmetric framework. However, apparently
Jauslin's treatment is not so directly linked to Witten's approach. To
provide a more Witten-like picture to this discrete case was one of
the motivations for our work.
The paper is organized as follows. A discussion of the
supersymmetric discrete master operator is presented in the next
section, where we introduce a discrete master second derivative and its
``square root" that enables us to write down a formal Riccati equation. In
this way we come close to Witten's treatment allowing us to solve the
simple case of homogeneous RWs.
In Sec. III one can find an outline of Jauslin's ``addition of
eigenvalues" method. Then, in Sec. IV we make a simple generalization
to two uncorrelated axes for homogeneous RW's, and we end up with some
conclusions.

\section*{II. SUSY Master Operator: Formalism}

The 1D master equation used in I is the common one referring to three neighbor
sites (or states)
\begin{equation}
-\frac{\partial P(n,t)}{\partial t}= -f(n-1)P(n-1,t)-b(n+1)P(n+1,t)+
[f(n)+b(n)]P(n,t)= (M_{-}P)(n,t),
\end{equation}
where $f(n-1)$ is the transition rate for one-step forward jumps starting
at the $(n-1)$ site and $b(n+1)$ is the corresponding rate for the backward
one-step jumps starting at the $(n+1)$ site. The jumps are random and the
site (or state) is following a Markovian process with the evolution of the
probability given by the master equation.

It is well known that the solution
of the time-independent equation $M_{-}P_{st}=0$ is \cite{K1}
\begin{equation}
P_{st}(n)=const\times \prod _{j} \frac{f(j-1)}{b(j)}.
\end{equation}
This form is a result of the detailed balancing condition
$f(n)P_{st}(n)=b(n+1)P_{st}(n+1)$.
The master operator $M_{-}$ can be made Hermitian by defining a new
function
\begin{equation}
\psi (n,t)=[P_{st}(n)]^{-1/2}P(n,t).
\end{equation}
This function satisfies an operatorial equation of the type
$ -\frac{\partial \psi (n,t)}{\partial t}=(H_{-}\psi )(n,t)$, which is
similar to Eq. (1),
$$
-\frac{\partial \psi (n,t)}{\partial t}=-[f(n-1)b(n)]^{1/2}\psi (n-1,t)
-[f(n)b(n+1)]^{1/2}\psi (n+1,t)+[f(n)+b(n)]\psi (n,t).  \eqno(4)   $$
The ``Hamiltonian" operator $H_{-}$ is a symmetric positive operator with
respect to the appropriate discrete $l_{2}$ scalar product. Also, for a
normalizable $P_{st}$ the lowest eigenvalue of $H_{-}$ is nought and
the ground state eigenfunction is $\phi _{gr,-} = [P_{st}]^{1/2}$.
The factorizing operators have been found in I to be
$$
(A^{+}\psi)(n)=b^{1/2}(n+1)\psi (n+1) - f^{1/2}(n)\psi (n)
   \eqno(5)  $$
and
$$(A^{-}\psi)(n)=b^{1/2}(n)\psi (n-1) - f^{1/2}(n)\psi (n).  \eqno(6)  $$

These two factorizing operators can be written as
$$A^{+} = b^{1/2}(n+1)\partial _{n}^{+} + [b^{1/2}(n+1)-f^{1/2}(n)]=
b^{1/2}(n+1)\partial _{n}^{+} +W^{+}(n)  \eqno(7a)  $$
or
$$A^+\psi (n)=b^{1/2} _{n+1}\left ( \partial ^{+}_{n}
+\frac{W_n ^{+}}{b^{1/2} _{n+1}}\right )\psi (n)    \eqno(7b)  $$
and
$$A^{-}=-b^{1/2}(n)\partial _{n-1}^{+} +  [b^{1/2}(n)-f^{1/2}(n)]=
-b^{1/2}(n)\partial _{n-1}^{+} +W^{-}(n)   \eqno(8a)   $$
or
$$A^-\psi (n)=b^{1/2}_{n}\left (-\partial ^{+}_{n-1}
+\frac{W_n ^{-}}{b^{1/2} _{n}}\right )\psi (n),    \eqno(8b)  $$
where $\partial _{n}^{+}\psi (n)=\psi (n+1)-\psi (n)$ and
$\partial _{n-1}^{+}\psi (n)=\psi (n)-\psi (n-1)$ are discrete derivative
operations, and the functions
$W_{-}(n)$ and $W_{+}(n)$ correspond to  the superpotential of the
continuous limit as applied to $\psi (n)$.
In this way, $H_{-}=A^{+}A^{-}$, and the superpartner will be
$H_{+}=A^{-}A^{+}$. Following I, we suppose the
``Hamiltonian" $H_{+}$ to be of the same type as $H_{-}$. That means
solutions of the type $b_+(n)=f_-(n)$ and $f_+(n)=b_-(n+1)$.

We write now the matrix form, i.e., the nilpotent master supercharges,
of the factorizing operators,
$ Q_M ^{-}= A_-\sigma _+ =
\left( \begin{array}{cc}
0 & 0 \\
A^{-} & 0
\end{array} \right)$
and
$Q_M ^{+}= A_+\sigma _-  =
\left( \begin{array}{cc}
0 & A^{+} \\
0 & 0
\end{array} \right)$;
$\sigma _-= \left( \begin{array}{cc}
0 & 1 \\
0 & 0
\end{array} \right)$ and
$\sigma _+=\left( \begin{array}{cc}
0 & 0 \\
1 & 0
\end{array} \right)$ are Pauli matrices.
In this realization, the matrix form of the
``Hamiltonian" operator is
$$ {\cal H} =
\left( \begin{array}{cc}
A^+ A^- & 0 \\
0 & A^- A^+
\end{array} \right) =
\left( \begin{array}{cc}
H_- & 0 \\
0 & H_+
\end{array} \right)
\eqno(9)  $$
defining the partner ``Hamiltonians" as diagonal elements of the matrix one.
They are partners in the sense that they are isospectral, apart from
the ground state $\phi _{gr,-}$ of $H_-$, which is not included in the
spectrum of $H_+$.

There is also one way of writing a formal discrete Witten scheme
for the master operator. Firstly we write the operator action in the form
$$H_-\psi (n)=-(b_n f_n)^{1/2}
\left\{\Bigg[\left(\frac{b_{n+1}}{b_n}\right)^{1/2}\psi _{n+1} -2\psi _n
+\left(\frac{f_{n-1}}{f_n}\right)^{1/2}\psi _{n-1}\Bigg]+
\Bigg[2-\frac{b_n+ f_n}{(b_n f_n)^{1/2}}\Bigg]\psi _n \right\}
\eqno(10) $$
and consider this operator as a one-site (local) operator, i.e., acting on
$\psi (n)$.
For this, one should introduce a discrete second derivative
operation of the type
$$\partial ^{2}_{M,n}\psi (n)=
\Bigg[\left(\frac{b_{n+1}}{b_n}\right)^{1/2}\psi _{n+1} -2\psi _n
+\left(\frac{f_{n-1}}{f_n}\right)^{1/2}\psi _{n-1}\Bigg]  \eqno(11) $$
that we call a {\em master discrete second derivative operation}.
In this way the master equation can be written ``locally" as follows:
$$H_-\psi (n)= (b_n f_n)^{\frac{1}{2}} \Bigg[-\partial ^{2}_{M,n} +
\left(\frac{b_n+f_n}{(b_n  f_n)^{1/2}}-2\right)\Bigg]\psi (n).  \eqno(12)   $$
Then one can proceed formally with the Witten scheme, by defining the
``square root" operator of $\partial ^{2}_{M,n}$, or more exactly we have to go
from the second master derivative to the first one by square-rooting,
an operation to which
we do not give here any rigorous meaning and we just denote it as
$\sqrt{\partial ^{2}_{M,n}}$. Then one may  consider the symmetric discrete
part $S(n)=\frac{1}{2}[\frac{b_n+ f_n}{(b_n f_n)^{1/2}}-2]$ as playing the role
of the ``Schr\"odinger" potential that we consider as a given quantity.
Consequently
$$H_-=2(b_nf_n)^{\frac{1}{2}}[{\cal A}^+{\cal A}^- +\epsilon],   \eqno(13)$$
where we introduced in the usual way the symmetry breaking parameter
$\epsilon$ (factorization energy),
and
$${\cal A}^{\pm}=\frac{1}{\sqrt{2}}
[\pm \sqrt{\partial ^{2}_{M,n}} + W_M(n)].  \eqno(14) $$
The formal Riccati equation for the master superpotential $W_M (n)$ would be
$$W_M ^2(n)+\sqrt{\partial ^{2}_{M,n}}W_M(n)= 2[S(n)-\epsilon].  \eqno(15) $$
The master superpotential $W_M (n)$ appears to be
a factor acting on $\psi (n)$, but taking into account the nonlocal
(three-site)
character of the discrete master second derivative the above Riccati equation
is formal as far as we introduced merely notations in order to put into
evidence the similarity with the standard supersymmetric (susy) quantum
mechanics.

An explicit case that can be solved completely is that of jump rates
independent of the location along the axis, i.e., $b_n=c_1,\;f_n=c_2$, with
$c_1$ and $c_2$ two positive constants.
For free RW's $c_1=c_2$, while for $c_1\neq c_2$ we have an
anisotropic RW. One can also normalize to unity, $c_1 + c_2=c=1$.
In these cases there is a similarity between the master discrete second
derivative and the popular
discrete second derivative obtained when one starts from the continuum and
makes the usual discretization up to $O(\Delta ^2)$ terms,
$$f''(x)=\frac{1}{\Delta ^2}[f(x+\Delta)-2f(x)+f(x-\Delta)]+ O(\Delta ^{2}).
\eqno(16)  $$
Therefore in the continuum limit we identify the master second derivative
with a common spatial second derivative implying the evolution equation
$$H_-\psi (n)=\Bigg[-(c_1 c_2)^{\frac{1}{2}}
\frac{\partial ^2}{\partial n^2}+
[(c_1 +c_2)-2(c_1 c_2)^{\frac{1}{2}}]\Bigg]\psi (n)    \eqno(17) $$
which being a normal Schr\"odinger equation is easily manipulated within
the Witten susy scheme. Indeed, by the rescaling
$x=2^{-1/2}(c_1 c_2)^{-1/4} n$ we get $H_- \psi (x)=
[-\frac{1}{2} \frac{\partial ^{2}}{\partial x^{2}} + S(c_1,c_2)]\psi (x)$.
This is a ``Schr\"odinger-type" equation for a unit mass particle in the
$S(c_1,c_2)$ potential.
Next, in the framework of the addition of variables method (see the next
section) we have to solve
the equation $H_-\psi=-\lambda \psi$, with $\lambda$ a positive constant,
corresponding to the case $\epsilon=-\lambda$, less than the ground-state
energy \cite{Suk85}.
In the normalized to unity case, $c_1+c_2=1$, the stationary solutions must
be of the following hyperbolic type:
$$\psi _{0} (x) \propto \exp \left\{
\pm \sqrt{2[1-2(c_1 c_2)^{1/2}+\lambda]}x\right\}.
\eqno(18) $$
It is really easy to derive the exact form of $\psi _{0}$ if we require
the ground-state wave function to be an
acceptable one, that is, $l_2$ ``integrable" and satisfying correct boundary
conditions. The latter ones are that it must vanish at the end points
$\pm \infty$, in agreement with the Sturm-Liouville theory (see I).
To be more concrete, let us consider the free RW, $c_1=c_2=1/2$. Then
$S(1/2,1/2)=0$ and the corresponding susy quantum mechanical problem is that
of a free particle.
The Riccati master equation, Eq. (15), can be written
$$W^2(x)+\frac{\partial W}{\partial x}=2\lambda    \eqno(19) $$
with the solution $W(x)=\sqrt{2\lambda}\tanh [\sqrt{2\lambda}(x-x_0)]$, where
$x_0$ is an integration constant. Then in the unbroken susy case it is known
that
$$\psi _{0} =\exp \Bigg[-\int  W(x')dx'\Bigg].   \eqno(20) $$
By introducing the $\tanh$ solution one gets immediately
$\psi _0=\cosh(\sqrt{2\lambda}x)$, i.e., similar to the findings of Jauslin.

\section*{III. The Addition of Eigenvalues}
Jauslin's method of addition of eigenvalues is clearly exposed in I and we make
here a brief outline. In fact, this method is a disguised form of a well-known
procedure in susy quantum mechanics \cite{Suk85}, \cite{Ro1}, namely the
factorization energy $\epsilon$ less than the ground-state energy.
The idea consists in constructing a {\em shifted} ``Hamiltonian" with respect
to an initial one, $H_0$. Both Hamiltonians have zero energies for their
ground states, but all the other eigenvalues of the shifted one are
displaced with a chosen arbitrary distance, say $\lambda _1$, from the
ground-state energy. Then, the proposal of Jauslin is to identify the
shifted Hamiltonian with $H_{+}$. This condition leads to the following
system of equations for the new jump rates $f_1$ and $b_1$:
$$ f_1(n)b_1(n)=\alpha ^{2}(n) = f_0(n)b_0(n-1),  \eqno(21)  $$
$$ f_1(n+1)+b_1(n)=\beta (n)= f_0(n) + b_0(n)+ \lambda _1.  \eqno(22) $$
The following ansatz,
$$f_1(n)=\alpha (n)\frac{\phi _{0} (n)}{\phi _{0} (n-1)},  \eqno(23a)  $$
$$b_1(n)=\alpha (n)\frac{\phi _{0}(n-1)}{\phi _{0}(n)},   \eqno(23b) $$
causes the geometrical mean equation, the first part of Eq. (21), to be
identically satisfied,
and moreover, Eq. (22) is turned into a Schr\"odinger-type equation for the
initial Hamiltonian at the eigenvalue $-\lambda _1$,
$H_0\phi _{0}=-\lambda _{1}\phi _{0}$.

The real importance of the above method comes out when one is going to
add even more eigenvalues (i.e., continuing the shifting)
by an iteration procedure of the form
$$H_{k-1}\phi _{k-1}=-\lambda _k\phi _{k-1}\;, k\geq 1.   \eqno(24)  $$
For the second eigenvalue, generating bistability in the stationary
probability, one can obtain easily
$$f_2 (n)=[f_1 (n)b_1 (n-1)]^{1/2}\frac{\phi _1(n)}{\phi _1(n-1)},
\eqno(25a) $$
$$b_2 (n)=[f_1 (n)b_1 (n-1)]^{1/2}\frac{\phi _1 (n-1)}{\phi _1 (n)}.
\eqno(25b)  $$
The corresponding stationary probability, which is the quantity of interest, is
$$P_2^{st} (n)=const\times
\frac{[f_1 (n)b_1 (n-1)]^{-1/2}}{\phi _1 (n)\phi _1 (n-1)}.  \eqno(26)  $$
In Eq. (26), the jump rates $f_1$ and $b_1$ are determined through Eqs. (23)
above, while the function $\phi _1$ is given by
$$\phi _1(n)=b^{1/2} _{1}\stackrel{-}{\phi} _{0}(n;\lambda _1+\lambda _2)-
f^{1/2} _{1}(n)\stackrel{-}{\phi} _{0}(n-1;\lambda _1+\lambda _2) \eqno(27)  $$
As sketched in I, the expression for the function $\stackrel{-}{\phi} _{0}$
comes out from the Sturm-Liouville theory. When the second
eigenvalue is made some two orders of magnitude smaller than the first one,
a well-defined
bistability (or bifurcation) in the stationary transition probability starts
developing. This was shown in I for the simple case of an initial free random
walk, i.e., $f_0(n)=b_0(n)=1/2$, and will be shown here for the two-axes
generalization (Figs. 1 and 2).

\section*{IV. Uncorrelated Two-Axes Generalization}

The simplest generalization of the results obtained in I is to the case of
homogeneous (i.e., free and/or anisotropic)
RW's along two uncorrelated discrete axes.
When the two axes are supposed uncorrelated, one can write down easily the
matrix form for the independent RW's along the two axes,
since this case corresponds to the separable 2D potentials in supersymmetric
quantum mechanics. The algebra of the supercharge operators can
be realized by writing $Q^-=q^-\times \sigma _+$ and $Q^+=q^+\times \sigma _-$,
with $q^-=\left( \begin{array}{cc}
A^- & 0 \\
0 & B^- \end{array} \right ) $  and
$q^+ =\left (\begin{array}{cc}
A^+ & 0 \\
0 & B^+ \end{array} \right ) $. The $\sigma _{\pm}$ matrices are again the
Pauli matrices. The symbols $A$ and $B$ correspond to the
first, and second axis, respectively, and are given by expressions of the type
(7a) and (7,b) or (8a) and (8,b) above.
The total 4$\times$4 ``Hamiltonian" matrix can be written as

$${\cal H}_{AB}=\pmatrix
{A^+A^- & 0 & 0 & 0 \cr
0 & B^+B^- & 0 & 0 \cr
0 & 0 & A^-A^+ & 0 \cr
0 & 0 & 0 & B^-B^+\cr}.
\eqno(28) $$
The supersymmetric partner ``Hamiltonians" are diagonalized 2$\times$2
matrices, with each diagonal component depending on one axis alone.

Even this trivial two-axes generalization implies nevertheless a richer
spectrum of possibilities for the physical situations, allowing us to
have various combinations for the two stationary states $P_2 ^{st} (n)$
and $P_2 ^{st} (m)$ on the two axes depending on the choice of the
two eigenvalues on each axis. We plotted for illustration a discrete free
RW case with bistability on one axis alone and another one with bistability on
both axes (see Figs. 1 and 2). The plotted functions are of the type
obtained by Jauslin $P_2 ^{st}= \frac{2}{\phi _1(n)\phi _1(n-1)}$ with
$$\phi _1(n)=\Bigg[\frac{1}{2a(n)}\Bigg]^{1/2}
\sinh(\gamma _{2}n-\delta _2)-
\Bigg[\frac{a(n)}{2}\Bigg]^{1/2}\sinh[\gamma _2(n-1)-\delta _2],
\eqno(29)  $$
where
$a(n)=\frac{\cosh(\gamma _1 n-\delta _1)}{\cosh[\gamma _1(n-1)-\delta _1]}$,
and $\gamma _1=$ arccosh$(1+\lambda _1)$,
$\gamma _2=$ arccosh$(1+\lambda _1 +\lambda _2)$, with $\lambda  _1$ and
$\lambda _2$ shifting constants, and $\delta _1$ and
$\delta _2$ arbitrary constants.

For homogeneous RW's in the continuous limit the matrices $q^{\pm}$ can
be written
in terms of the ``superpotentials" $W_A(n)$ and $W_B(m)$ on the two axes
in the form
$$q^{\pm}=\pmatrix
{\pm \frac{\partial}{\partial n} +W_A(n) & 0 \cr
0 & \pm \frac{\partial}{\partial m} + W_B(m) \cr}
 \eqno(30)$$
or
$$q^{\pm}=\pmatrix
{\pm \frac{\partial}{\partial x}+\sqrt{\Delta}\tanh [\sqrt{\Delta} (x- x_0)] &
0 \cr
0 & \pm \frac{\partial}{\partial y} +
\sqrt{\Delta}\tanh [\sqrt{\Delta} (y -y_0)] \cr},
   \eqno(31)$$
where $\Delta =2[1-2(c_1c_2)^{1/2}+\lambda]$, and we rescaled the coordinates
$x=2^{-1/2}(c_1c_2)^{-1/4}n$,
$y=2^{-1/2}(c_1c_2)^{-1/4}m$; $x_0$ and $y_0$ are integration constants.
The explicit form of $W_{A,B}$ results from solving a Riccati equation of the
type Eq. (19) with $\lambda$ replaced by $[1-2(c_1c_2)^{1/2}+\lambda]$.

\section*{V. Conclusions}

The simple supersymmetric algebraic schemes for RW's developed in
this paper, as a continuation of I, may
have many applications, both for the 1D case and even more for the
two-axes case of the preceding section, either uncorrelated or
correlated. It is known that the three-sites (or states) master equation
applies to one-step processes \cite{H4} such as  the number of molecules
in a chemical species,
the number of photons in a lasing mode, or the number of electrons on a
capacitor. It will be of interest to apply supersymmetric algebraic schemes
to multistep processes as well.
It will be also very useful to find closed analytical forms for the
more realistic correlated two-axes case, which must be studied carefully.

%%%%%%%%%%%%%%%%%%%%%%%%%%%%%  END OF THE PAPER   %%%%%%%%%%%%%%%%%%%%%%%%%

\section*{Acknowledgments}
This work was partially supported by the CONACYT Grant No. F246-E9207.
M.R. was supported by a CONACYT Graduate Fellowship, and wishes to thank
Mauro Napsuciale for enlightening conversations.

%%%%%%%%%%%%%%%%%%%%%%%%%  REFERENCES   %%%%%%%%%%%%%%%%%%%%%%%%%%%%%%%%

\begin{thebibliography}{99}
\bibitem[*]{byline} Electronic address: rosu@ifug.ugto.mx
\bibitem[*]{byline} Electronic address: mareyes@fnalv.fnal.gov
\bibitem{BB}
         M. Bernstein and L.S. Brown, Phys. Rev. Lett. {\bf 52}, 1933 (1984);
         F. Marchesoni, P. Sodano, and M. Zannetti, {\em ibid}. {\bf 61},
         1143 (1988).
\bibitem{W1}
         E. Witten, Nucl. Phys. B {\bf 185}, 513 (1981).

\bibitem{J1}
         H. R. Jauslin, Phys. Rev. A {\bf 41}, 3407 (1990) (paper I).

\bibitem{K1}
         N.G. van Kampen, {\it Stochastic Processes in Physics and Chemistry}
         (North-Holland, Amsterdam, 1981).

\bibitem{Suk85}
         C. V. Sukumar, J. Phys. A {\bf 18}, 2917 (1985).

\bibitem{Ro1}
         J. L. Rosner, Ann. Phys. {\bf 200}, 101 (1990).

\bibitem{H4}
         P. Hanggi {\em et al}., Phys. Rev. A {\bf 29}, 371 (1984).

\newpage
{\bf Figure captions}
\bigskip

FIG. 1.  Stationary states $P_2^{st}(n)$ and $P_2^{st}(m)$ for two uncorrelated
free RW along two axes A
and B with the two eigenvalues as follows:
(a) $\lambda _{1A}= 0.01$ and $\lambda _{2A}=0.01$;
b) $\lambda _{1B}=0.01$ and $\lambda _{2B}=0.0005$.
This is a case with bistability along one axis only.
\bigskip

FIG. 2.  The same as in Fig. 1 but with the following set of eigenvalues:
(a) $\lambda _{1A}=0.01$ and $\lambda _{2A}=0.005$;
(b) $\lambda _{1B}=0.01$ and $\lambda _{2B}=0.0002$.
This case displays bistability along both axes.

\end{document}